\documentclass[article,twocolumn,superscriptaddress]{revtex4}
\usepackage[dvips,final]{graphicx}

\begin{document}

\newcommand{\ethaffil}{Laboratory of Physical Chemistry, ETH Zurich, CH-8093 Zurich, Switzerland}
\newcommand{\humbaffil}{Nano-Optics, Humboldt University, Hausvogteiplatz 5-7, D-10117 Berlin,
Germany}
\newcommand{\brasilaffil}{Departamento de F\'{\i}sica, Universidade Federal de
Pernambuco, 50670-901 Recife-PE, Brazil}

\title{Controlled coupling of counterpropagating whispering-gallery modes by a single Rayleigh scatterer: a classical problem in a quantum optical light}

\author{A.~Mazzei}
\affiliation{\humbaffil}
\author{S.~G\"{o}tzinger}
\affiliation{\ethaffil}
\author{L.~de S.~Menezes}
\affiliation{\brasilaffil}
\author{G.~Zumofen}
\affiliation{\ethaffil}
\author{O.~Benson}\email{oliver.benson@physik.hu-berlin.de}
\affiliation{\humbaffil}
\author{V.~Sandoghdar}\email{vahid.sandoghdar@ethz.ch}
\affiliation{\ethaffil}

\begin{abstract}
We present experiments where a single subwavelength scatterer is
used to examine and control the back-scattering induced coupling
between counterpropagating high-Q modes of a microsphere resonator.
Our measurements reveal the standing wave character of the resulting
symmetric and antisymmetric eigenmodes, their unbalanced intensity
distributions, and the coherent nature of their coupling. We discuss
our findings and the underlying classical physics in the framework
common to quantum optics and provide a particularly intuitive
explanation of the central processes.
\end{abstract}

\maketitle

The radiative properties of atoms can be strongly modified by
coupling them to resonators~\cite{Berman}. A historical corner stone
of this field of research, known as Cavity Quantum Electrodynamics
(CQED), was set in 1946 by E.~M. Purcell who proposed that the
radiation rate of an oscillating dipole at wavelength $\lambda$ can
be enhanced by a factor $F=3Q\lambda^3/4\pi^2 V_m$ in a resonant
cavity of quality factor $Q$ and mode volume $V_m$~\cite{Berman}.
This so-called Purcell effect holds in the dissipative \emph{weak}
coupling regime where the cavity finesse is small so that the atomic
radiation remains dominated by its coupling to the bath of the
electromagnetic modes. In the \emph{strong} coupling regime,
coherent exchange of energy between the atom and the resonator
causes the atomic resonance to lose its identity and to become
replaced by a doublet. These phenomena have been studied for more
than three decades~\cite{Domokos:00,Klinner:06,Aoki:06,Hennessy:07}
although the \emph{in-situ} manipulation of a single emitter in a
single mode of a high-Q microresonator remains a
challenge~\cite{Aoki:06,Hennessy:07}. In this Letter, we consider
the controlled coupling of a classical nano-object to a high-finesse
whispering-gallery mode (WGM) microresonator. We discuss both
theoretically and experimentally the resulting coherent coupling
between two degenerate counterpropagating WGMs and the modification
of the Rayleigh scattering rate. Our findings show that the concepts
of the strong and weak coupling play a central role even in this
fully classical system.

\begin{figure}[bht]
\includegraphics[width=0.48\textwidth]{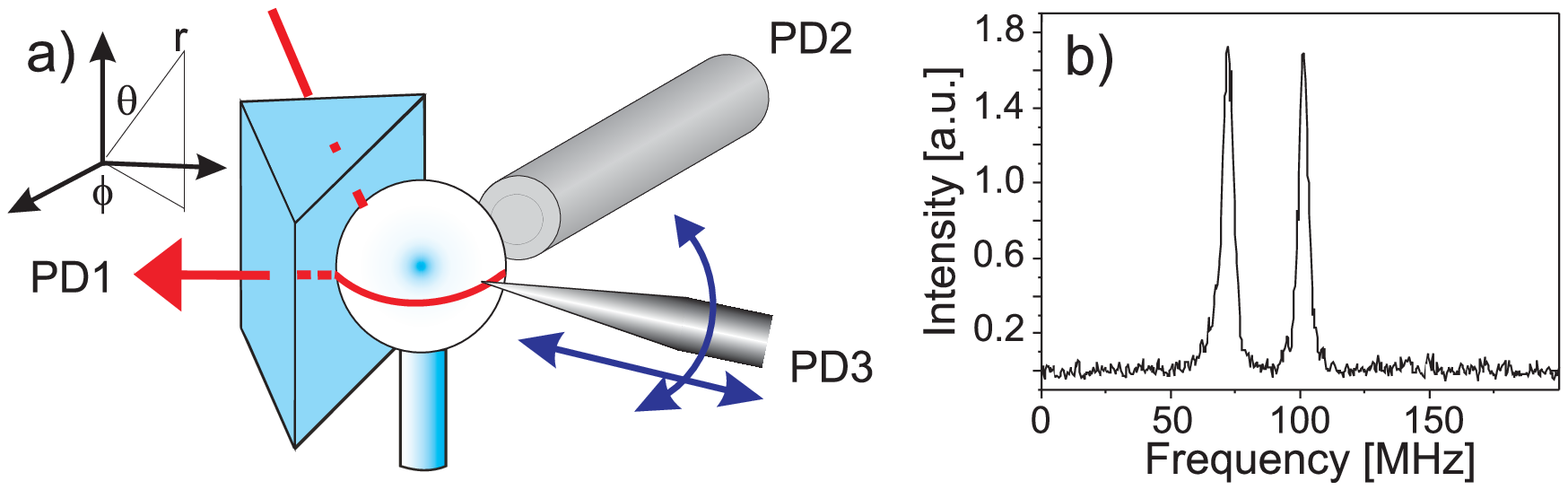}
\caption{a) Whispering gallery modes of a microsphere are excited
via a prism. A glass fiber tip can be positioned in ($r,
\theta,\phi$) close to the sphere surface. The resonator spectrum
can be recorded on PD1 in transmission, on PD2 from global
scattering out of the sphere via a multimode fiber, and on PD3
through the fiber tip. b) An example of the doublet spectrum
recorded on PD2.} \label{setup}
\end{figure}

The resonators in our work consist of microspheres melted at the end
of silica fibers~\cite{Braginsky:89}. Such spheres support very
high-Q WGMs and have been studied by several
groups~\cite{Collot-93,Kippenberg:02,Gorodetsky:00}. About ten years
ago, it was discovered that the high-Q resonances of these cavities
are often composed of doublets~\cite{Weiss-95OL}. Such a mode
splitting has been since discussed in conjunction with various WGM
resonators~\cite{Krupka:96,Gorodetsky:00,Kippenberg:02,Borselli:05,Srinivasan:07}.
It turns out that mode splitting has been observed in other ring
resonators and has been explained as the result of the coupling
between the electric fields $E_{c}$ and $E_{cc}$ of the degenerate
clockwise (c) and counter clockwise (cc) modes via back scattering.
The new superpositions states (+) and ($-$) are described by
\begin{equation}
E_{+}= a E_{c}+ b E_{cc}~;~ E_{-}= a E_{c}- b E_{cc}. \label{fields}
\end{equation}
Here $a$ and $b$ are complex coefficients. In the simplest case, the
coupling between $E_c$ and $E_{cc}$ can be caused by a
reflector~\cite{Spreeuw:91,Venugopalan:93}. In the case of WGM
resonators, however, it has been suggested that backscattering from
a distribution of residual subwavelength inhomogeneities in the
glass matrix or on its surface is the source of this
coupling~\cite{Weiss-95OL,Kippenberg:02,Gorodetsky:00}. The orders
of magnitude of the doublet splitting can be correctly estimated
from classical electrodynamic considerations following this
hypothesis~\cite{Weiss-95OL,Gorodetsky:00,Kippenberg:02,Borselli:05,Bourgeois:05,Srinivasan:07}.
Nevertheless, the direct link between the spectral features of a
doublet and the nanoscopic details of the backscattering sources has
not been demonstrated experimentally, and a proper treatment of the
losses inflicted by the scatterers is missing in the literature. In
fact, an intuitively perplexing and interesting question arises in
this context: given that the radiation of a subwavelength scatterer
is nearly isotropic and that the angle subtended by a typical cavity
mode is merely about $10^{-4}$~rad~\cite{Goetzinger:06}, how could
the rate of scattering back into a cavity mode dominate the rate of
scattering out of the resonator to ensure the population of $E_{+}$
and $E_{-}$?

The schematics of our experimental arrangement is shown in
Fig.~\ref{setup}a. A narrow-band diode laser ($\lambda=670$~nm,
linewidth $<300$~kHz, tuning range $\approx 60$~GHz) was used to
excite the WGMs via a prism. Photodetector PD1 was used to record
spectra in transmission through the prism whereas PD2 captured the
light globally scattered out of the microsphere into a multimode
fiber. Figure~\ref{setup}b displays a typical doublet with a
splitting of $29$~MHz and $Q\simeq 8 \times 10^7$ recorded on PD2.
The peaks represent the intensities $|E_{+}|^2$ and $|E_{-}|^2$ of
the symmetric and antisymmetric eigenmodes. Following the procedure
described in Ref.~\cite{Mazzei:05}, we applied scanning near-field
optical microscopy (SNOM) to map the spatial intensity distribution
of the WGMs on PD3 and to identify the fundamental mode of the
resonator, which exhibits a single intensity maximum in the $r$ and
$\theta$ directions. In our previous works, we have shown that an
uncoated glass tip might broaden and shift cavity modes depending on
their $Q$ and on the tip size~\cite{Goetzinger-02OL}. Here we
demonstrate that a subwavelength tip can modify or induce the
coupling between the degenerate c and cc microsphere modes.

\begin{figure}[bht]
\includegraphics[width=0.45\textwidth]{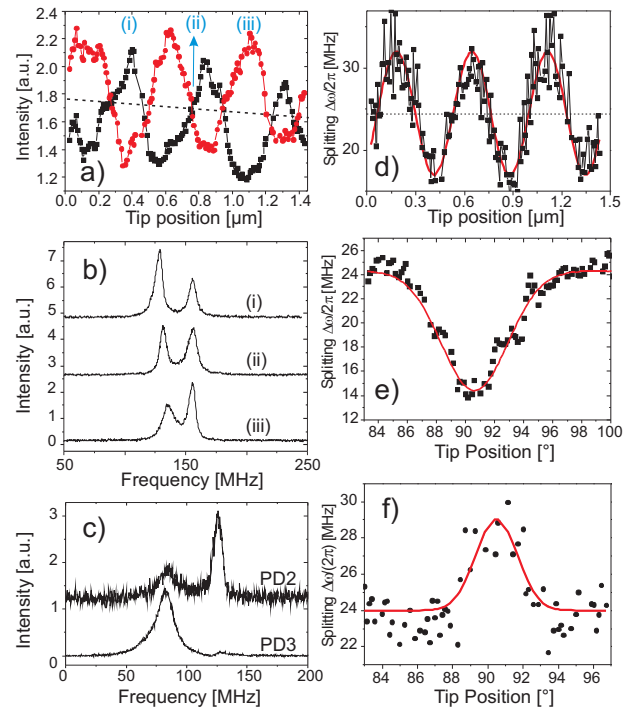}
\caption{a) The intensities of the two peaks in a doublet as a
function of the tip position along the equator. The dotted line
marks the slight intensity drift in the detection. b) Spectra
recorded at positions (i-iii) in Fig.~(a). The spectra are displaced
vertically for clarity. c) Simultaneously measured spectra on PD2
and PD3. d) The recorded splitting corresponding to the data in
Fig.~(a). The solid curve shows a sinusoidal fit. e) and f) The
variations of the mode splitting as a function of the tip in the
$\theta$ direction for positions (i) and (iii), respectively. The
solid curves display fits according to the spatial mode function of
the fundamental WGM.} \label{standingwave}
\end{figure}

Equation~(\ref{fields}) implies that the interference between the c
and cc running modes should give rise to sine and cosine standing
waves along the equator. The locations of the nodes and antinodes of
$E_+$ and $E_-$ are automatically established by the random
distribution of a large number of inhomogeneities in the silica
sphere~\cite{Weiss-95OL,Gorodetsky:00}. To visualize this effect, we
have scanned a sharp fiber tip along the equator (i.e. in the $\phi$
direction) of an already split fundamental WGM and have recorded
spectra at each point (note that the radial coordinate of the tip is
kept constant using a shear-force feedback~\cite{Mazzei:05}).
Figure~\ref{standingwave}a shows that the intensities of the two
peaks of a doublet undergo out of phase periodic modulations as a
function of the tip location. A slight slope of the middle base line
is attributed to a drift in the shear-force tip-sphere
stabilization. At location (i), the tip is positioned in the node of
the symmetric mode and the antinode of the antisymmetric mode. Thus,
it induces loss in the $E_+$ mode while it leaves $E_-$ nearly
unaffected. Position (iii) shows the opposite counterpart of (i)
whereas at position (ii) both modes are affected equally strongly.

As shown in Fig.~\ref{standingwave}b, the three spectra reveal that
in addition to a change in the intensity balance of the doublets,
their splittings are also modified. Figure~\ref{standingwave}d
displays the modulation of the splitting about its initial value of
24~MHz shown by the dotted line. Interestingly, we find that the tip
can not only increase the mode splitting, but it can also decrease
it. This is due to a destructive interference between the field
scattered by the tip and the field scattered by the inhomogeneities
in the microsphere that gave rise to the initial splitting.
Figures~\ref{standingwave}e and f provide further data on the
increase and decrease of the mode splitting as the tip was scanned
in the $\theta$ direction for two different $\phi$ positions spaced
by half of the interference period along the equator. Finally,
Fig.~\ref{standingwave}c plots the resonance spectra recorded
simultaneously on PD2 and PD3, i.e. via global scattering from the
sphere and via the fiber tip. The different lineshapes on the two
channels might seem unexpected at first. However, this effect shows
that if the tip is placed in an antinode of $E_-$ or $E_+$, it
efficiently extracts photons out of that mode, leading to a larger
signal in the fiber tip and thus a lower intensity in the cavity
mode. On the contrary, the mode that is less perturbed is stronger
in the resonator and is nearly uncoupled to the fiber tip. It is
evident that the mode that is coupled to the tip has experienced an
additional broadening.

We now show that the radiation properties of a subwavelength object
such as its scattering rate are modified much in the same manner as
those of the spontaneous emission of an atom. Our guiding thought is
that many central features of CQED, including the modification of
the mode density in a resonator, can be traced to the spatial
character of the modes and should be thus shared by classical cavity
electrodynamics. We first present a simple treatment of the
\emph{free-space} Rayleigh scattering using a semi-quantum
electrodynamic (semi-QED) approach, where the material scatterer is
treated classically while the field is quantized. Then we will
discuss the modification of the scattering rate when the scatterer
is coupled to a resonator.

Let us assume that a freely propagating photon is incident on a
subwavelength spherical scatterer of radius $a$ and refractive index
$n$. We take the photon to be in a mode $\widehat{E}_k$ with volume
$V_k$, frequency $\omega_k$ and a linear polarization along the unit
vector $\mathbf{\epsilon}_k$ such that $\widehat{\mathbf{E}}_k={\cal
E}_k\mathbf{\epsilon}_k(\widehat{a}^\dag_k+\widehat{a}_k)$ where
$\widehat{a}^\dag_k$ and $\widehat{a}_k$ are the usual creation and
annihilation operators and  ${\cal E}_k=\sqrt{\hbar
\omega_k/2\epsilon_0 V_k}$. In the limit where the scatterer is
considerably smaller than $\lambda$, it can be described by a
dipolar polarizability $\alpha$~\cite{Jackson-book} so that the
induced dipole moment operator reads
$\widehat{\mathbf{p}}_k=\varepsilon_0\alpha {\cal E}_k
(\widehat{a}^\dag_k+\widehat{a}_k) \mathbf{\epsilon}_k$. Thus the
interaction energy between this dipole moment and another mode
$\widehat{E}_j$ becomes,
\begin{equation}
\widehat{\cal{V}}_{k,j} = -\widehat{\mathbf{p}}_k\cdot
\widehat{\mathbf{E}}_j= \hbar g_{kj}
(\widehat{a}^{\dag}_k\widehat{a}_j+\widehat{a}^{\dag}_j\widehat{a}_k)
\label{Coupling-energy}
\end{equation}if we neglect the terms that do not conserve photon numbers. Here we have taken $g_{kj}=-\alpha \sqrt{\omega_k \omega_j} |\mathbf{\epsilon}^\ast_j \cdot \mathbf{\epsilon}_k|/ 2\sqrt{V_k V_{vac}}$ and have set $V_j=V_{vac}$ for all vacuum modes $j$ into which the incident beam is scattered. The system Hamiltonian
becomes~\cite{EPAPS}
\begin{equation}
\widehat{H}=\hbar \omega_k
\widehat{a}^\dag_k\widehat{a}_k+\sum_j\hbar \omega_j
\widehat{a}^\dag_j\widehat{a}_j+ \sum_j\hbar
g_{kj}(\widehat{a}^\dag_k\widehat{a}_j+\widehat{a}^\dag_j\widehat{a}_k),
\label{hamiltonian}
\end{equation} leading to the Heisenberg equation of motion
\begin{equation}
i\dot{\widehat{a}}_k = \omega_k \widehat{a}_k+\sum_j
g_{kj}\widehat{a}_j. \label{heisenberg}
\end{equation}
The last term in Eq.~(\ref{heisenberg}) signifies the scattering of
the incident field into all vacuum modes. Following the
Weisskopf-Wigner formalism~\cite{Milonni-book}, we find the rate
\begin{equation}
\Gamma_{R}=\frac{2\omega_k^2 V_{vac}}{3\pi c^3} g_R^2=\frac{\alpha^2
\omega_k^2}{6 \pi c^3 V_{k}} \label{Gamma-Rayleigh-free}
\end{equation}for this scattering event~\cite{EPAPS}. Here we have
restricted ourselves to $\omega_k=\omega_j$ for elastic scattering
and have used the notation $g_R= -\alpha \omega_k/ 2 \sqrt{V_k
V_{vac}}$. Now we can calculate the Rayleigh scattering cross
section $\sigma_R$~\cite{Jackson-book} by considering the total
power radiated by the scatterer according to $I_{\rm
inc}\sigma_R=\hbar \omega_k \Gamma_R$. Given that $I_{\rm inc}=\hbar
\omega_kc/V_k$ and $\alpha=4\pi a^3|\frac{n^2-1}{n^2+2}|$, one
obtains the well-known relation\begin{equation} \sigma_R=\frac{8\pi
k^4 a^6}{3} \left|\frac{n^2-1}{n^2+2}\right|^2.
\label{Rayleigh-cross-sect}
\end{equation}

We note that a rigorous quantum optical treatment of scattering is
not frequently discussed in the literature~\cite{Dalton:99} and goes
beyond the scope of this paper. However, the fact that $\sigma_R$
can be derived via the Weisskopf-Wigner formalism using quantized
fields provides a robust support for the intuitive expectation that
a modification of the mode density, for example in a resonator or in
front of a mirror, could also lead to a change in the Rayleigh
scattering rate. The corresponding Purcell effect offers a physical
explanation for the question posed earlier. The rate with which
energy is transferred from $E_c$ to $E_{cc}$ is enhanced by $F$ and
is given by $\eta F$ where $\eta$ is the geometric factor
determining the fraction of the solid angle subtended by the mode.
Therefore, a fundamental WGM with $V_m\simeq130\,\mu m^3$ (sphere
diameter $30\, \mu m$) and $Q=10^8$ yields $F\sim 10^4$,
compensating for the very small geometric acceptance of the order of
$10^{-4}$. The influence of the Purcell factor $F$ also explains why
reducing the cavity $Q$ results in the disappearance of light in the
counterpropagating mode as reported previously~\cite{Weiss-95OL}.

\begin{figure}[h]
\includegraphics[width=0.47\textwidth]{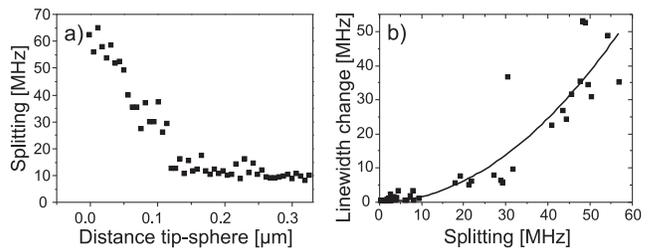}
\caption{a) The mode splitting measured as a function of sphere-tip
separation in the \emph{r} direction. b) Plot of the measured
splitting versus the additional tip-induced broadening. The solid
curve displays the fit to a quadratic function.}
\label{QvsSplitting}
\end{figure}

Having shown that the density of states plays a central role in the
description of Rayleigh scattering, next we consider the coupling of
two counterpropagating \emph{cavity} modes $E_c$ and $E_{cc}$ via a
single Rayleigh scatterer. The details of the calculations are
provided in the supplementary materials of this paper~\cite{EPAPS}.
Returning to a classical notation, we find
\begin{equation}
\begin{array}{ll}
 \dot {\tilde E}_- &=\left(  -i \Delta -  \gamma_0 \right)  \tilde E_- + \kappa_0 \\
  \dot {\tilde E}_+ &= \left(-i \Delta + 2 i g -   \gamma_0 - \Gamma \right)
    \tilde E_+  +\kappa_0.
 \end{array}\label{strong-coupling}
\end{equation}Here we have defined ${\widetilde{E}}=e^{i\omega t} E$, $\Delta=\omega -\omega_c$ shows the detuning of the laser frequency $\omega$, $\kappa=\kappa_0 e^{i\omega t}$ is the mode excitation rate, and $2\gamma_0$ denotes the
unperturbed cavity linewidth. When dealing with Rayleigh scattering
out of a cavity, we have to take into account the spatial variation
$f(\mathbf {r})$ of  $E_c$ and $E_{cc}$ in the resonator mode. Going
back to the definition of $g_{kj}$ and noting that $V_j=V_k$ is the
WGM volume $V_m$ of the two modes, we thus obtain~\cite{EPAPS}
\begin{equation}
\begin{array}{ll}
2g=-\alpha f^2(\mathbf {r}) \omega_c/ V_m \\
\Gamma=\alpha^2f^2(\mathbf {r}) \omega_c^4/6\pi c^3V_m
\end{array}\label{g-gamma}
\end{equation} for the mode splitting and broadening, respectively.

A close scrutiny of Eq.~(\ref{strong-coupling}) shows that if $|2g|$
is sufficiently large to overcome $\gamma_0$ and if
$2|\frac{n^2-1}{n^2+2}|<(\frac{\lambda}{2\pi a})^3$ to assure that
$|2g|>\Gamma$, a mode splitting is resolved. This is similar to the
case of the strong coupling in CQED where the coherent exchange of
energy between the cavity mode and an atom leads to a mode splitting
if the coupling coefficient $|g|$ becomes larger than the cavity
linewidth. As in CQED, the mode splitting $|2g|$ grows with
decreasing $V_m$ and increasing $\omega$, but the atomic dipole
strength is replaced here by $\alpha$, and the splitting is
asymmetric. The antisymmetric mode remains unperturbed (see
Eq.~(\ref{strong-coupling})) whereas the symmetric mode undergoes an
additional broadening given by $\Gamma$. This is a consequence of
the fact that the phases of the new eigenmodes are self-adjusted so
that the local scatterer is placed in a node (antinode) of the
antisymmetric (symmetric) mode. This unbalanced splitting has been
recently also observed for the coupling of a far-detuned cold atom
ensemble to a high-Q ring cavity~\cite{Klinner:06}. Indeed, in that
case several million atoms detuned from their transition resonances
also behave essentially as a large dielectric object.

Equations~(\ref{g-gamma}) predict a linear dependence between the
tip-induced splitting and line broadening if $f(\mathbf{r})$ is
varied. However, due to the finite spatial extent of the tip and the
WGMs, $\alpha$ effectively grows as the tip enters the mode. This
position dependence leads then to a quadratic relationship between
the tip-induced broadening and splitting. To investigate this
effect, we have moved the tip in the radial direction and have
recorded spectra at each location. Figure~\ref{QvsSplitting}a shows
that as expected, the mode splitting becomes larger when the tip
enters the evanescent field of the microsphere.
Figure~\ref{QvsSplitting}b plots the tip-induced splitting versus
the increase in the linewidth of $E_+$, confirming their quadratic
dependence.

\begin{figure}[h]
\includegraphics[width=0.47\textwidth]{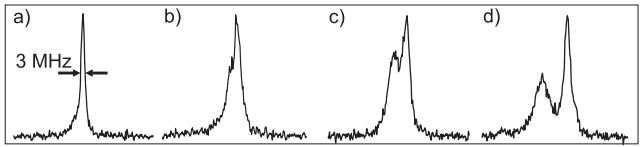}
\caption{a-d) WGM spectra recorded at different $\theta$ values of
the tip location. At position a) the tip is nearly outside the
spatial profile of the mode whereas at position d) it is in the mode
maximum, i.e. the sphere equator.} \label{weak2strong}
\end{figure}

To realize an ideal scenario for studying the interaction of a
single well-defined scatterer with the fundamental mode of a
microsphere, we have searched for spheres in which no mode splitting
was observable in the beginning and it was created only when the tip
was introduced. Figure~\ref{weak2strong} presents four spectra of a
resonance as the tip was scanned in the $\theta$-direction from
outside the mode (a) to the maximum of the mode at the equator (d)
at constant separation from the sphere surface. The $E_+$ mode is
shifted in frequency by 13 MHz and broadened by about 6 MHz. Given
that $|2g|/\Gamma=3\lambda^3/4\pi^2\alpha$ according to
Eqs.~(\ref{g-gamma}), the observed ratio of the splitting to
broadening implies a radius of $a\sim140$~nm for a spherical
Rayleigh scatterer. This is in very good quantitative agreement with
the experimental parameters, considering a typical value of
50-100~nm for the radii of curvature of SNOM tips and accounting for
the overlap of the conical tip taper with the evanescent part of the
mode.

In conclusion, we have considered the phenomenon of Rayleigh
scattering both in free space and in the presence of a resonator. By
using a semi-QED approach, we have pointed out the roles of the
modification of the density of states and of the Purcell effect in
classical scattering. Our results demonstrate that although the
introduction of a scatterer into a high-finesse resonator might be
commonly thought to introduce losses, it can mediate a coherent
coupling of the resonator modes and cause their consequent normal
mode splitting.

\section*{Acknowledgments} We are grateful for financial support from the Deutsche Forschungsgemeinschaft SP1113, the Swiss National
Foundation (ETH group), NaF\"{o}G, Berlin (A. M.) and the Alexander
von Humboldt Stiftung (L. de S. M.).

\newpage


\begin{thebibliography}{99}
\expandafter\ifx\csname
natexlab\endcsname\relax\def\natexlab#1{#1}\fi
\expandafter\ifx\csname bibnamefont\endcsname\relax
  \def\bibnamefont#1{#1}\fi
\expandafter\ifx\csname bibfnamefont\endcsname\relax
  \def\bibfnamefont#1{#1}\fi
\expandafter\ifx\csname citenamefont\endcsname\relax
  \def\citenamefont#1{#1}\fi
\expandafter\ifx\csname url\endcsname\relax
  \def\url#1{\texttt{#1}}\fi
\expandafter\ifx\csname urlprefix\endcsname\relax\def\urlprefix{URL
}\fi \providecommand{\bibinfo}[2]{#2}
\providecommand{\eprint}[2][]{\url{#2}}

\bibitem[{\citenamefont{Berman}(1994)}]{Berman}
\bibinfo{author}{\bibfnamefont{P.~R.} \bibnamefont{Berman}},
  \emph{\bibinfo{title}{Cavity Quantum Electrodynamics}}
  (\bibinfo{publisher}{Academic Press}, \bibinfo{year}{1994}).

\bibitem[{\citenamefont{Domokos et~al.}(2000)\citenamefont{Domokos, Gangl, and
  Ritsch}}]{Domokos:00}
\bibinfo{author}{\bibfnamefont{P.}~\bibnamefont{Domokos}},
  \bibinfo{author}{\bibfnamefont{M.}~\bibnamefont{Gangl}}, \bibnamefont{and}
  \bibinfo{author}{\bibfnamefont{H.}~\bibnamefont{Ritsch}},
  \bibinfo{journal}{Opt. Commun.} \textbf{\bibinfo{volume}{185}},
  \bibinfo{pages}{115} (\bibinfo{year}{2000}).

\bibitem[{\citenamefont{Klinner et~al.}(2006)\citenamefont{Klinner, Lindholdt,
  Nagorny, and Hemmerich}}]{Klinner:06}
\bibinfo{author}{\bibfnamefont{J.}~\bibnamefont{Klinner}},
  \bibinfo{author}{\bibfnamefont{M.}~\bibnamefont{Lindholdt}},
  \bibinfo{author}{\bibfnamefont{B.}~\bibnamefont{Nagorny}}, \bibnamefont{and}
  \bibinfo{author}{\bibfnamefont{A.}~\bibnamefont{Hemmerich}},
  \bibinfo{journal}{Phys. Rev. Lett.} \textbf{\bibinfo{volume}{96}},
  \bibinfo{pages}{023002} (\bibinfo{year}{2006}).

\bibitem[{\citenamefont{Aoki et~al.}(2006)\citenamefont{Aoki, Dayan, Wilcut,
  Bowen, Parkins, Kippenberg, Vahala, and Kiimble}}]{Aoki:06}
\bibinfo{author}{\bibfnamefont{T.}~\bibnamefont{Aoki}, et al.},
\bibinfo{journal}{Nature} \textbf{\bibinfo{volume}{443}},
\bibinfo{pages}{671} (\bibinfo{year}{2006}).

\bibitem[{\citenamefont{Hennessy et~al.}(2007)\citenamefont{Hennessy, Badolato,
  Winger, Gerace, Atature, Gulde, Falt, Hu, and Imamoglu}}]{Hennessy:07}
\bibinfo{author}{\bibfnamefont{K.}~\bibnamefont{Hennessy}, et al.},
  \bibinfo{journal}{Nature} \textbf{\bibinfo{volume}{445}},
  \bibinfo{pages}{896} (\bibinfo{year}{2007}).

\bibitem[{\citenamefont{Braginsky et~al.}(1989)\citenamefont{Braginsky,
  Gorodetsky, and Ilchenko}}]{Braginsky:89}
\bibinfo{author}{\bibfnamefont{V.~B.} \bibnamefont{Braginsky}},
  \bibinfo{author}{\bibfnamefont{M.~L.} \bibnamefont{Gorodetsky}},
  \bibnamefont{and} \bibinfo{author}{\bibfnamefont{V.~S.}
  \bibnamefont{Ilchenko}}, \bibinfo{journal}{Phys. Lett. A}
  \textbf{\bibinfo{volume}{137}}, \bibinfo{pages}{393} (\bibinfo{year}{1989}).

\bibitem[{\citenamefont{Collot et~al.}(1993)\citenamefont{Collot,
  Lefevre-Seguin, Brune, Raimond, and Haroche}}]{Collot-93}
\bibinfo{author}{\bibfnamefont{L.}~\bibnamefont{Collot}},
  \bibinfo{author}{\bibfnamefont{V.}~\bibnamefont{Lefevre-Seguin}},
  \bibinfo{author}{\bibfnamefont{M.}~\bibnamefont{Brune}},
  \bibinfo{author}{\bibfnamefont{J.}~\bibnamefont{Raimond}}, \bibnamefont{and}
  \bibinfo{author}{\bibfnamefont{S.}~\bibnamefont{Haroche}},
  \bibinfo{journal}{Eur. Phys. Lett.} \textbf{\bibinfo{volume}{23}},
  \bibinfo{pages}{327} (\bibinfo{year}{1993}).

\bibitem[{\citenamefont{Kippenberg et~al.}(2002)\citenamefont{Kippenberg,
  Spillane, and Vahala}}]{Kippenberg:02}
\bibinfo{author}{\bibfnamefont{T.~J.} \bibnamefont{Kippenberg}},
  \bibinfo{author}{\bibfnamefont{S.~M.} \bibnamefont{Spillane}},
  \bibnamefont{and} \bibinfo{author}{\bibfnamefont{K.~J.}
  \bibnamefont{Vahala}}, \bibinfo{journal}{Opt. Lett.}
  \textbf{\bibinfo{volume}{27}}, \bibinfo{pages}{1669} (\bibinfo{year}{2002}).

\bibitem[{\citenamefont{Gorodetsky et~al.}(2000)\citenamefont{Gorodetsky,
  Pryamikov, and Ilchenko}}]{Gorodetsky:00}
\bibinfo{author}{\bibfnamefont{M.~L.} \bibnamefont{Gorodetsky}},
  \bibinfo{author}{\bibfnamefont{A.~D.} \bibnamefont{Pryamikov}},
  \bibnamefont{and} \bibinfo{author}{\bibfnamefont{V.~S.}
  \bibnamefont{Ilchenko}}, \bibinfo{journal}{J. Opt. Soc. Am. B}
  \textbf{\bibinfo{volume}{17}}, \bibinfo{pages}{1051} (\bibinfo{year}{2000}).

\bibitem[{\citenamefont{Weiss et~al.}(1995)\citenamefont{Weiss, Sandoghdar,
  Hare, Lefevre-Seguin, Raimond, and Haroche}}]{Weiss-95OL}
\bibinfo{author}{\bibfnamefont{D.~S.} \bibnamefont{Weiss}, et al.},
  \bibinfo{journal}{Opt. Lett.} \textbf{\bibinfo{volume}{20}},
  \bibinfo{pages}{1835} (\bibinfo{year}{1995}).

\bibitem[{\citenamefont{Krupka et~al.}(1996)\citenamefont{Krupka, Blondy, Cros,
  Guillon, and Geyer}}]{Krupka:96}
\bibinfo{author}{\bibfnamefont{J.}~\bibnamefont{Krupka}},
  \bibinfo{author}{\bibfnamefont{P.}~\bibnamefont{Blondy}},
  \bibinfo{author}{\bibfnamefont{D.}~\bibnamefont{Cros}},
  \bibinfo{author}{\bibfnamefont{P.}~\bibnamefont{Guillon}}, \bibnamefont{and}
  \bibinfo{author}{\bibfnamefont{R.~G.} \bibnamefont{Geyer}},
  \bibinfo{journal}{IEEE Trans. Micro. Theo. Tech.}
  \textbf{\bibinfo{volume}{44}}, \bibinfo{pages}{1097} (\bibinfo{year}{1996}).

\bibitem[{\citenamefont{Borselli et~al.}(2005)\citenamefont{Borselli, Johnson,
  and Painter}}]{Borselli:05}
\bibinfo{author}{\bibfnamefont{M.}~\bibnamefont{Borselli}},
  \bibinfo{author}{\bibfnamefont{T.~J.} \bibnamefont{Johnson}},
  \bibnamefont{and} \bibinfo{author}{\bibfnamefont{O.}~\bibnamefont{Painter}},
  \bibinfo{journal}{Opt. Exp.} \textbf{\bibinfo{volume}{13}},
  \bibinfo{pages}{1515} (\bibinfo{year}{2005}).

\bibitem[{\citenamefont{Srinivasan and Painter}(2007)}]{Srinivasan:07}
\bibinfo{author}{\bibfnamefont{K.}~\bibnamefont{Srinivasan}} \bibnamefont{and}
  \bibinfo{author}{\bibfnamefont{O.}~\bibnamefont{Painter}},
  \bibinfo{journal}{Phys. Rev. A} \textbf{\bibinfo{volume}{55}},
  \bibinfo{pages}{023814} (\bibinfo{year}{2007}).

\bibitem[{\citenamefont{Spreeuw and Woerdman}(1991)}]{Spreeuw:91}
\bibinfo{author}{\bibfnamefont{R.~J.~C.} \bibnamefont{Spreeuw}}
  \bibnamefont{and} \bibinfo{author}{\bibfnamefont{J.~P.}
  \bibnamefont{Woerdman}}, \bibinfo{journal}{Physica B}
  \textbf{\bibinfo{volume}{75}}, \bibinfo{pages}{96} (\bibinfo{year}{1991}).

\bibitem[{\citenamefont{Venugopalan et~al.}(1993)\citenamefont{Venugopalan,
  Kumar, and Ghosh}}]{Venugopalan:93}
\bibinfo{author}{\bibfnamefont{A.}~\bibnamefont{Venugopalan}},
  \bibinfo{author}{\bibfnamefont{D.}~\bibnamefont{Kumar}}, \bibnamefont{and}
  \bibinfo{author}{\bibfnamefont{R.}~\bibnamefont{Ghosh}},
  \bibinfo{journal}{Pramana J. Phys.} \textbf{\bibinfo{volume}{40}},
  \bibinfo{pages}{107} (\bibinfo{year}{1993}).

\bibitem[{\citenamefont{Bourgeois and Giordano}(2005)}]{Bourgeois:05}
\bibinfo{author}{\bibfnamefont{P.-Y.} \bibnamefont{Bourgeois}}
  \bibnamefont{and} \bibinfo{author}{\bibfnamefont{V.}~\bibnamefont{Giordano}},
  \bibinfo{journal}{IEEE Trans. Microw. Theo. Tech.}
  \textbf{\bibinfo{volume}{53}}, \bibinfo{pages}{3185} (\bibinfo{year}{2005}).

\bibitem[{\citenamefont{G\"otzinger et~al.}(2006)\citenamefont{G\"otzinger,
  de~S.~Menezes, Mazzei, K\"uhn, Sandoghdar, and Benson}}]{Goetzinger:06}
\bibinfo{author}{\bibfnamefont{S.}~\bibnamefont{G\"otzinger}, et al.},
  \bibinfo{journal}{Nano Lett.} \textbf{\bibinfo{volume}{6}},
  \bibinfo{pages}{1151} (\bibinfo{year}{2006}).

\bibitem[{\citenamefont{Mazzei et~al.}(2005)\citenamefont{Mazzei,
  G\"{o}tzinger, de~S.~Menezes, Sandoghdar, and Benson}}]{Mazzei:05}
\bibinfo{author}{\bibfnamefont{A.}~\bibnamefont{Mazzei}},
  \bibinfo{author}{\bibfnamefont{S.}~\bibnamefont{G\"{o}tzinger}},
  \bibinfo{author}{\bibfnamefont{L.}~\bibnamefont{de~S.~Menezes}},
  \bibinfo{author}{\bibfnamefont{V.}~\bibnamefont{Sandoghdar}},
  \bibnamefont{and} \bibinfo{author}{\bibfnamefont{O.}~\bibnamefont{Benson}},
  \bibinfo{journal}{Opt. Comm.} \textbf{\bibinfo{volume}{250}},
  \bibinfo{pages}{428} (\bibinfo{year}{2005}).

\bibitem[{\citenamefont{G\"otzinger et~al.}(2002)\citenamefont{G\"otzinger,
  Benson, and Sandoghdar}}]{Goetzinger-02OL}
\bibinfo{author}{\bibfnamefont{S.}~\bibnamefont{G\"otzinger}},
  \bibinfo{author}{\bibfnamefont{O.}~\bibnamefont{Benson}}, \bibnamefont{and}
  \bibinfo{author}{\bibfnamefont{V.}~\bibnamefont{Sandoghdar}},
  \bibinfo{journal}{Opt. Lett.} \textbf{\bibinfo{volume}{27}},
  \bibinfo{pages}{80} (\bibinfo{year}{2002}).

\bibitem[{\citenamefont{Jackson}(1999)}]{Jackson-book}
\bibinfo{author}{\bibfnamefont{D.}~\bibnamefont{Jackson}},
  \emph{\bibinfo{title}{Classical Electrodynamics}} (\bibinfo{publisher}{Wiley
  and Sons}, \bibinfo{year}{1999}).


\bibitem[{EPAPS()}]{EPAPS}
\bibinfo{note}{The EPAPS Supplementary material presents the details of the calculations leading to equations (\ref{Gamma-Rayleigh-free}) and (\ref{strong-coupling}).}


\bibitem[{\citenamefont{Milonni}(1994)}]{Milonni-book}
\bibinfo{author}{\bibfnamefont{P.~W.} \bibnamefont{Milonni}},
  \emph{\bibinfo{title}{The Quantum Vacuum}} (\bibinfo{publisher}{Academic
  Press, New York}, \bibinfo{year}{1994}).


\bibitem[{\citenamefont{Dalton et~al.}(1999)\citenamefont{Dalton, Barnette, and
  Knight}}]{Dalton:99}
\bibinfo{author}{\bibfnamefont{B.~J.} \bibnamefont{Dalton}},
  \bibinfo{author}{\bibfnamefont{S.~M.} \bibnamefont{Barnette}},
  \bibnamefont{and} \bibinfo{author}{\bibfnamefont{P.~L.}
  \bibnamefont{Knight}}, \bibinfo{journal}{J. Mod. Opt.}
  \textbf{\bibinfo{volume}{46}}, \bibinfo{pages}{1107} (\bibinfo{year}{1999}).

\end{thebibliography}
\end{document}